\documentclass[fleqn,11pt,twoside]{article}

\usepackage{amsthm,amsthm,amssymb, color, xcolor,epsfig, graphics, subfigure}

\usepackage{amsmath, graphicx, latexsym, lscape }
\usepackage[colorlinks=true, linkcolor=black, citecolor=black, urlcolor=black]{hyperref}

\newcommand{\figref}[1]{\figurename~\ref{#1}}
\makeatletter
\newcommand{\copyrightnote}[2]{{\renewcommand{\thefootnote}{}
 \footnotetext{\small\it
\begin{flushleft}
 \copyright \ #1   #2  
\end{flushleft}}}}

\newcommand{\Name}[1]{\begin{flushleft}
                       \LARGE \bf #1
                       \end{flushleft}\vspace{-3mm}}

\newcommand{\Author}[1]{\begin{flushleft}
                       \it #1 \end{flushleft}}

\newcommand{\Address}[1]{\begin{flushleft}
                       \it #1 \end{flushleft}}

\newcommand{\Date}[1]{\begin{flushleft}
                      \small  \it #1 \end{flushleft}}

\newcommand{\evenhead}{Author \ name}
\newcommand{\oddhead}{Article \ name}

\renewcommand{\@evenhead}{
\hspace*{-3pt}\raisebox{-15pt}[\headheight][0pt]{\vbox{\hbox to \textwidth
{\thepage \hfil \evenhead}\vskip4pt \hrule}}}
\renewcommand{\@oddhead}{
\hspace*{-3pt}\raisebox{-15pt}[\headheight][0pt]{\vbox{\hbox to \textwidth
{\oddhead \hfil \thepage}\vskip4pt\hrule}}}
\renewcommand{\@evenfoot}{}
\renewcommand{\@oddfoot}{}

\long\def\@makecaption#1#2{%
  \vskip\abovecaptionskip
  \sbox\@tempboxa{\small \textbf{#1.}\ \ #2}%
  \ifdim \wd\@tempboxa >\hsize
    {\small \textbf{#1.}\ \ #2}\par
  \else
    \global \@minipagefalse
    \hb@xt@\hsize{\hfil\box\@tempboxa\hfil}%
  \fi
  \vskip\belowcaptionskip}

\newcommand{\JNMPnumberwithin}[3][\arabic]{%
  \@ifundefined{c@#2}{\@nocounterr{#2}}{%
    \@ifundefined{c@#3}{\@nocnterr{#3}}{%
      \@addtoreset{#2}{#3}%
      \@xp\xdef\csname the#2\endcsname{%
        \@xp\@nx\csname the#3\endcsname .\@nx#1{#2}}}}%
}

\newcommand{\resetfootnoterule} {
  \renewcommand\footnoterule{%
  \kern-3\p@
  \hrule\@width.4\columnwidth
  \kern2.6\p@}
}

\renewcommand{\footnoterule}{}

\makeatother

\theoremstyle{definition}

\begin{document}

\renewcommand{\evenhead}{ {\LARGE\textcolor{blue!10!black!40!green}{{\sf \ \ \ ]ocnmp[}}}\strut\hfill 
D Latini
}
\renewcommand{\oddhead}{ {\LARGE\textcolor{blue!10!black!40!green}{{\sf ]ocnmp[}}}\ \ \ \ \   
Rigid Body Rotors in Planar Potentials
}

\thispagestyle{empty}
\newcommand{\FistPageHead}[3]{
\begin{flushleft}
\raisebox{8mm}[0pt][0pt]
{\footnotesize \sf
\parbox{150mm}{{Open Communications in Nonlinear Mathematical Physics}\ \  \ {\LARGE\textcolor{blue!10!black!40!green}{]ocnmp[}}
\ \ Vol.6 (2026) pp
#2\hfill {\sc #3}}}\vspace{-13mm}
\end{flushleft}}

\FistPageHead{1}{\pageref{firstpage}--\pageref{lastpage}}{ \ \ Article}

\strut\hfill

\strut\hfill

\copyrightnote{The author(s). Distributed under a Creative Commons Attribution 4.0 International License}

\Name{Rigid Body Rotors in Planar Potentials: A Novel type of Superintegrable Mechanical Systems}

\Author{Danilo Latini$^{1,2}$}

\Address{$^1$ Universit\`a degli Studi di Milano, Dipartimento di Matematica ``Federigo Enriques", \\
	Via Cesare Saldini 50, 20133 Milano, Italy\\
	$^2$ INFN Sezione di Milano, Via Giovanni Celoria 16, 20133 Milano, Italy\\
	Email: danilo.latini@unimi.it}

\Date{Received July 6, 2026; Accepted November 25, 2026}

\setcounter{equation}{0}

\smallskip

\noindent
{\bf Citation format for this Article:}\newline
 Danilo Latini, 
 Rigid Body Rotors in Planar Potentials: A Novel type of Superintegrable Mechanical Systems,
{\it Open Commun. Nonlinear Math. Phys.}, {\bf 6}, ocnmp:18702, \pageref{firstpage}--\pageref{lastpage}, 2026.

\strut\hfill

\noindent
{\bf The permanent Digital Object Identifier (DOI) for this Article:}\newline
{\it 10.46298/ocnmp.18702}

\strut\hfill

\begin{abstract}
\noindent 
We investigate the superintegrability of rigid body rotors coupled to planar systems. In particular, we study the isotropic harmonic oscillator in two dimensions, with its (central) force acting on the rotor’s center of mass constrained to move in the plane. By including an internal rotational degree of freedom described by a rigid rotor, the resulting planar system possesses three degrees of freedom: two translational and one rotational. When the orbital motion and the internal rotation are tuned to resonance, additional integrals of motion arise, extending the hidden symmetry algebras of the underlying models. For the oscillator, the well-known $\mathfrak{su}(2)$ symmetry algebra can be enlarged by the presence of the rotor, with the conserved momentum $p_{\theta}$ reasonably playing the role of a deformation parameter. These algebraic structures remain to be properly understood, and we hope that this short work will serve as an invitation to further investigate these interesting models. To close the work, we also examine the oscillator in a vertical plane, in the presence of a rotor, under the effect of a uniform gravitational field, showing that the algebraic structure persists as a translated version of the isotropic case, as expected. In all these settings, the extended dynamics admits five functionally independent integrals, thereby confirming maximal superintegrability. Our simple yet nontrivial results suggest that rigid-body rotors provide a natural mechanism for generating new families of (resonant) superintegrable systems, along with their associated symmetry algebras, an outcome that aligns with the main objective of this work.
\end{abstract}

\label{firstpage}


\section{Introduction}
\label{intro}

\noindent Maximally Superintegrable systems \cite{MPW, BBM} in $N$ degrees of freedom represent a very special subset of finite-dimensional integrable Hamiltonian systems, being characterized by a total number of $2N-1$ functionally independent constants of motion, far more than the $N$ independent integrals required for Liouville integrability \cite{Liouville1853, whittaker}. Recall that, in the integrable case, the $N$ integrals must be in mutual involution. Moreover, under the usual regularity and compactness assumptions, the common level sets $F_i=\alpha_i$ $(i=1,\dots, N)$ are $N$-dimensional invariant tori, and the induced dynamics on each torus is quasi-periodic (conditionally periodic)~\cite{Arnol'd1997, Reshetikhin, Fasso}. Extremely important in the theory of superintegrability are those models that have integrals of motion quadratic in the momenta, as they are related to multiseparability in the Hamilton-Jacobi and Schr\"{o}dinger equations, respectively in classical and quantum mechanics \cite{MPW, 10.1088/978-0-7503-1314-8}. Building on the pioneering works \cite{FMSUW,WSUF,MSVW}, these systems have been classified in two and three dimensions on conformally flat spaces, and the corresponding program is currently being extended to higher dimensions $N$ (see \cite{KSV} and references therein). An important aspect about superintegrability is that special functions \cite{AS} arise as solutions of exactly solvable problems so, as a surprising consequence, the Askey scheme, which organizes hypergeometric orthogonal polynomials, emerges from contractions of superintegrable systems. As a matter of fact, every second-order superintegrable system in two dimensions can be obtained as an appropriate limiting case of a single “master” model: the generic three-parameter potential on the two-sphere, denoted S9 in the classification reported in \cite{MPW}. It follows that the associated quadratic symmetry algebras are likewise obtained by contraction from the quadratic algebra of 
S9. Concerning superintegrable systems in Flat Euclidean plane $E^2$ all models are known and shown to be exactly solvable together with their  hidden algebra (maximal parabolic subalgebra of
$\mathfrak{sl}(3, \mathbb{R})$) \cite{TTW}. They actually come from four different families, namely \cite{FMSUW}:
\begin{align}
	V_I(x,y)&=\alpha (x^2+y^2)+\frac{\beta}{x^2}+\frac{\gamma}{y^2} \, , \hskip 1.3 cm V_{II}(x,y)=\alpha (x^2+4y^2)+\frac{\beta}{x^2}+\gamma y \, ,\label{2Dnote1}\\
	V_{III}(x,y)&=\frac{\alpha}{r}+\frac{1}{r^2}\left(\frac{\beta}{\cos^2 (\varphi/2)} +\frac{\gamma}{\sin^2 (\varphi/2)} \right) \, ,\hskip 0.2cm V_{IV}(x,y)=\frac{\alpha}{r}+\frac{1}{\sqrt{r}}\bigl( \beta \cos( \varphi/2)\nonumber \\&\hskip 9.1cm+\gamma \sin( \varphi/2)\bigl) .
	\label{2Dnote2}
\end{align}

\noindent The classical trajectories, quantum energy levels, and wavefunctions of all these systems are known. Potentials $V_I$ and $V_{II}$ are deformations of the isotropic and anisotropic harmonic oscillator, respectively, while $V_{III}$ and $V_{IV}$ are deformations of the Kepler–Coulomb potential.

Superintegrable systems in $2D$, such as the harmonic oscillator and Kepler–Coulomb system, represent the classical prototypes of superintegrable systems, also taking into account their rather simple but still nontrivial symmetry algebra, namely $\mathfrak{su}(2)$ and $\mathfrak{so}(3)$ for bound states, respectively. Of course, they just appear as the $N=2$ case of the higher rank cases $\mathfrak{su}(N)$ and $\mathfrak{so}(N+1)$, which employ much more structure.

For two-dimensional (at least second-order) superintegrable systems, the classical trajectories, quantum spectra, and wavefunctions are by now essentially understood. In this work we therefore shift the focus to a different class of models inspired by the $2D$ setting but enriched by an additional rotational degree of freedom: a rigid-body rotor. The resulting system remains planar in its translational motion, and all forces involved act on the rigid body’s center of mass. However, coupling to an internal angular variable fundamentally changes the dynamics, transforming an otherwise purely translational $2D$ problem into a hybrid translational–rotational one.

\noindent A key role is now played by resonance conditions between the planar orbital motion and the rotor’s internal rotation. These resonances control the onset of additional conserved quantities and, consequently, the structure of the associated symmetry algebra.

\noindent This work is ongoing and is meant as a gentle invitation to the community to consider this class of superintegrable systems, which, to the best of our knowledge, has not yet been addressed in the existing literature. 

\noindent Here, we left open some of the main ideas to be addressed:
\begin{enumerate}
	
	\item Construct extended systems with three degrees of freedom (two translational plus one rotational).
	
	\item Analyze the resonance condition between orbital motion in the plane and the internal rotation.
	
	\item Exhibit the resulting integrals of motion and the associated symmetry algebras.
	
	\item Discuss possible applications, exact-solvability, and physical realizations (also in quantum mechanics).
	
\end{enumerate}

\noindent To illustrate some initial results along these lines, we focus on two very basic examples. Although simple, they are still nontrivial from the perspective we wish to emphasize in this paper. We divide the work into four  main sections:
\begin{itemize}
	\item In Section~\ref{sec2} we introduce and discuss a rigid-body rotor coupled to a planar system, which provides the central idea of the work.
	\item In Section~\ref{sec3} we apply the general framework to the harmonic potential, with particular emphasis on the counting of degrees of freedom and on the resonance condition leading to maximal superintegrability.
	\item In Section~\ref{sec4} we consider a closely related example, namely an oscillator in a vertical plane under gravity. This illustrates that even modest variations of the model can give rise to interesting and nontrivial features.
	\item In Section~\ref{sec5} we discuss on further works and open perspectives.
	
\end{itemize} 

\section{The Rigid Rotor Coupled to Planar Systems} 
\label{sec2}
\noindent Our initial point is a Mechanical system described with generalized coordinates\footnote{The generalization to $\mathbb{R}^N$ is straightforward.}:
\begin{equation}
	\boldsymbol{q}=\begin{pmatrix} q_1 \\ q_2 \\ q_3 \end{pmatrix} \in \mathbb{R}^3
\end{equation}
and we want to obtain the Hamiltonian:
\begin{equation}
	H(\boldsymbol{q},\boldsymbol{p})=T(\boldsymbol{q},\boldsymbol{p})+V(\boldsymbol{q}) \qquad \boldsymbol{q}\in \mathbb{R}^3 \quad \text{and} \quad \boldsymbol{p}\in \mathbb{R}^3
\end{equation}
starting from a Lagrangean:
\begin{equation}
	L(\boldsymbol{q},\dot{\boldsymbol{q}})=T(\boldsymbol{q},\dot{\boldsymbol{q}})-V(\boldsymbol{q})=\frac{1}{2}\dot{\boldsymbol{q}}^\text{T}A(\boldsymbol{q})\dot{\boldsymbol{q}}-V(\boldsymbol{q}) \, ,
\end{equation}
where $A(\boldsymbol{q}) \in \mathbb{R}^3 \times \mathbb{R}^3$ is symmetric and positive definite for each $\boldsymbol{q}$, which means that $\forall \, x \in \mathbb{R}^3$ then $\boldsymbol{q}^\text{T} A \boldsymbol{q}>0$ (each eigenvalue of $A$ is strightly positive). All the following analysis assumes the necessary regularity (derivability of $A$ and $V$). Now, for each $i=1,2,3$ we have: 
\begin{equation}
	p_i=\frac{\partial L}{\partial \dot{q}_i}=\sum_{j=1}^3A_{ij}(\boldsymbol{q})\dot{q}_j\equiv [A(\boldsymbol{q})\dot{\boldsymbol{q}}]_i
\end{equation}
and so: $\boldsymbol{p}=A(\boldsymbol{q})\dot{\boldsymbol{q}}$. Considering its inverse we obtain:
\begin{equation}
	\dot{\boldsymbol{q}}=A^{-1}(\boldsymbol{q})\boldsymbol{p}\, .
	\label{2.5}
\end{equation}
At this point, through Legendre transform: $H(\boldsymbol{q},\boldsymbol{p})=\boldsymbol{p}^\text{T} \dot{\boldsymbol{q}}-L(\boldsymbol{q},\dot{\boldsymbol{q}})$, with some simple algebraic manipulations and simplifications, we arrived to:
\begin{equation}
	H(\boldsymbol{q}, \boldsymbol{p})=\frac{1}{2}\boldsymbol{p}^{\text{T}}A^{-1}(\boldsymbol{q})\boldsymbol{p}+V(\boldsymbol{q})\, .
\end{equation}
The Hamiltonian we are interested in this work are of the type:

\begin{equation}
	\boldsymbol{q}=\begin{pmatrix} x\\ y \\ \theta \end{pmatrix} \in \mathbb{R}^2 \times S^1 \qquad 	\boldsymbol{p}=\begin{pmatrix} p_x \\ p_y \\ p_{\theta} \end{pmatrix} \in \mathbb{R}^3
\end{equation}

\begin{equation}
	A(\boldsymbol q):=\begin{pmatrix}
		M & 0 & 0\\
		0 & M & 0\\
		0 & 0 & I
	\end{pmatrix} \quad  \Rightarrow  \quad A^{-1}(\boldsymbol q):=
	\begin{pmatrix}
		1/M & 0 & 0\\
		0 & 1/M & 0\\
		0 & 0 & 1/I
	\end{pmatrix} \quad \text{and}\quad  V(\boldsymbol{q})=V(x,y) \, .
\end{equation}
And so, they have the following form:
\begin{equation}
	\frac12
	\begin{pmatrix} p_x & p_y & p_\theta \end{pmatrix}
	\begin{pmatrix}
		1/M & 0 & 0\\
		0 & 1/M & 0\\
		0 & 0 & 1/I
	\end{pmatrix}
	\begin{pmatrix} p_x \\ p_y \\ p_\theta \end{pmatrix}+V(x,y)\,,
\end{equation}

\noindent which translates to:
\begin{equation}
	H = \frac{p_x^2+p_y^2}{2M} + V(x,y) + \frac{p_\theta^2}{2I},
	\label{eq:ham} 
\end{equation}
where $I$ is the moment of inertia (with respect to the $z$-axes) of a homogeneous rigid body of mass $M$. 
This Hamiltonian describes the planar motion of a rigid body whose center of mass G, with coordinates $\boldsymbol{r_G}=(x,y)$, can move in the plane under the influence of a force $\boldsymbol{F}=-\boldsymbol{\nabla} V=-(V_x,  V_y)$, while the additional term $\tfrac{p_\theta^2}{2I}$ accounts for its rotational degree of freedom. The latter arises after Legendre transform from the term $\frac{1}{2} I \dot{\theta}^2$, with $(x,y,\theta)$ being generalized coordinates in the Lagrangean formalism. The associated Hamilton's equations read:
\begin{equation}
	\dot{x} = \frac{\partial H}{\partial p_x} = \frac{p_x}{M}, \qquad 
	\dot{y} = \frac{\partial H}{\partial p_y} = \frac{p_y}{M}, \qquad 
	\dot{\theta} = \frac{\partial H}{\partial p_\theta} = \frac{p_\theta}{I}, 
	\label{eq:q_dot}
\end{equation}
\begin{equation}
	\dot{p}_x = -\frac{\partial H}{\partial x} = -V_x, \qquad
	\dot{p}_y = -\frac{\partial H}{\partial y} = -V_y, \qquad
	\dot{p}_\theta = -\frac{\partial H}{\partial \theta} = 0 \, ,
	\label{eq:p_dot}
\end{equation}
which give rise to the following equations of motion:
\begin{align}
	M\ddot{x}+V_x=0\, , \qquad M\ddot{y}+V_y=0 \, ,\qquad  \ddot{\theta}=0 \, .
\end{align}
So, we see that the third equation immediately gives:

\begin{equation}
	\theta(t)=\theta_0+\frac{p_{\theta}}{I} t, \qquad \theta_0:=\theta(0) \, , \qquad p_{\theta}(0):=p_{\theta} \, , 
	\label{rot}
\end{equation}
which means we have uniform circular motion in the $\theta$ coordinate, meaning that the rigid body is rotating in the plane with constant angular velocity $\dot \theta:=p_{\theta}/I$. 
From this perspective, the choices of the initial conditions dramatically changes the physics of the problem, starting with zero angular velocity, for example, i.e. taking $p_{\theta}=0$ would collapse the Hamiltonian system to the standard 2D harmonic oscillator, as there is no more rotation (we dropped one degree of freedom). This is the case when the rigid body behaves like a particle of mass $M$, described by the Hamiltonian:
\begin{equation}
	H = \frac{p_x^2+p_y^2}{2M} + V(x,y),
	\label{eq:hamil} 
\end{equation}
that we know very well from the point of view of superintegrability (see \eqref{2Dnote1}-\eqref{2Dnote2}). We know, in particular, that in this case four second-order potentials appears, namely eqs. \eqref{2Dnote1}-\eqref{2Dnote2}  \cite{FMSUW}, keeping into account of course the formal replacement $M \to m$.

On the contrary, by taking $p_{\theta} \neq 0$ would result in a uniform rotation of the rigid body, providing in total two translational degrees of freedom plus a rotational one, namely three degrees of freedom.
We note that $\mathbb{R}^2 \times S^1$ and $\mathbb{R}^3$ are locally isometric. Therefore, to establish a connection with the classification results of Kalnins, Kress, and Miller (KKM), we can relate the matrix $A(\boldsymbol{q})$ to the metric tensor $g(\boldsymbol{q})$. Since $A(\boldsymbol{q})$ is a constant matrix, the corresponding metric is flat, and hence our systems fall within the geometric framework considered in the KKM classification. However, as we shall demonstrate below, the systems studied in this paper are not in general second-order superintegrable and therefore do not belong to the classes covered by the aforementioned classification. More precisely, in this work we investigate maximally superintegrable systems with three degrees of freedom, two translational and one rotational, possessing five functionally independent constants of motion, which include in general higher-order integrals in the momenta.
\begin{figure}[!ht]
	\centering
	\begin{tabular}{cc}
		\includegraphics[width=0.4\textwidth]{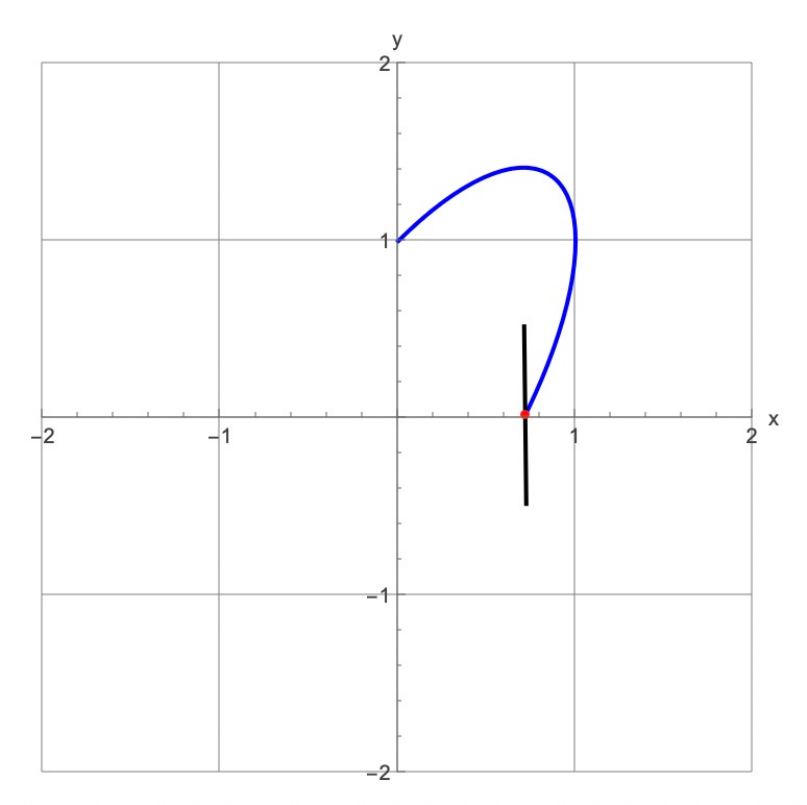} &
		\includegraphics[width=0.4\textwidth]{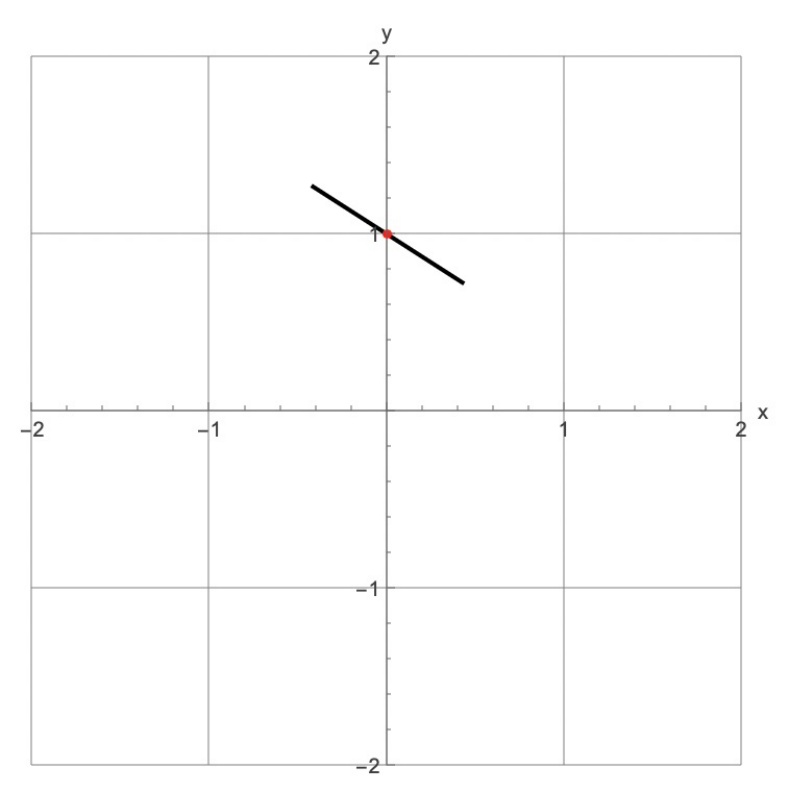} \\
		\vspace{2mm} & \vspace{2mm} \\
		\includegraphics[width=0.4\textwidth]{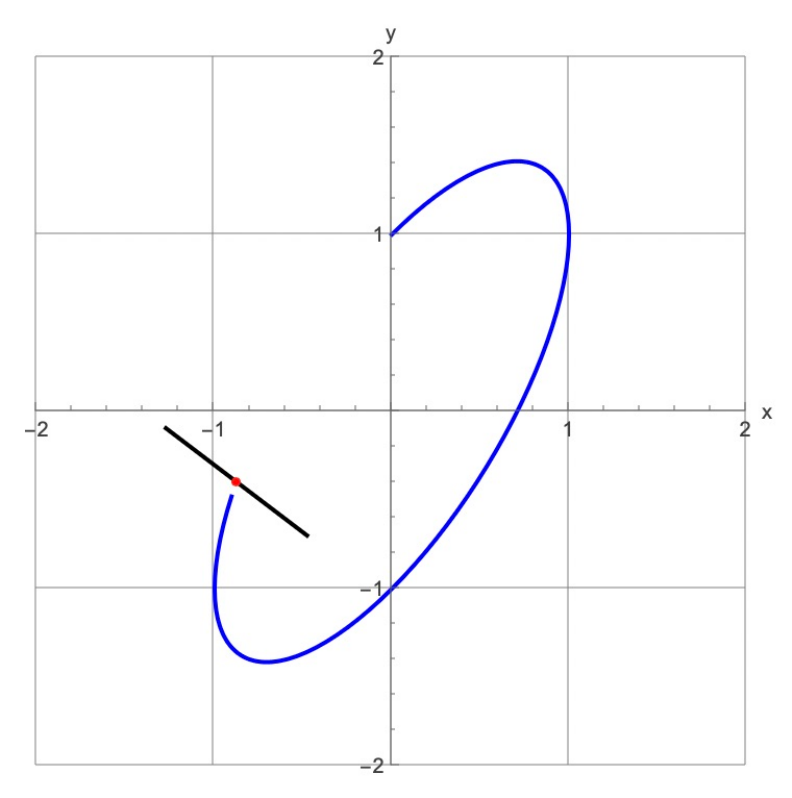} &
		\includegraphics[width=0.4\textwidth]{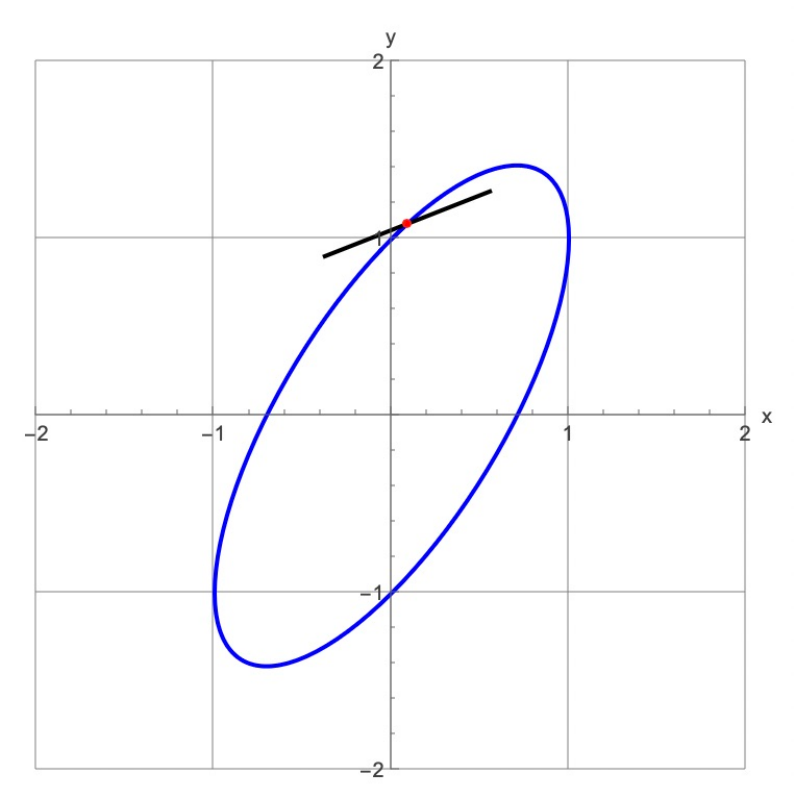}
	\end{tabular}
	\caption{A homogeneous rod of mass $M$ completes one full revolution over a single orbital cycle. The forces acting on the rotor’s center of mass are purely elastic, so the resulting trajectory is an ellipse, as expected.}
	\label{fig:grid}
\end{figure}

\noindent The first thing that we notice is that, for central potentials $V(x,y)=V(r)$, $r:=\sqrt{x^2+y^2}$ the Hamiltonian: 

\begin{equation}
	H = \frac{p_x^2+p_y^2}{2M} + V(r) + \frac{p_\theta^2}{2I},
	\label{eq:hamil} 
\end{equation}
is integrable, being endowed with the three constants of motion in involution:
\begin{equation}
	F_1= H, \quad F_2 = x p_y-y p_x \, , \quad F_3 = p_{\theta} \, .
	\label{eq:inte}
\end{equation}
Now, we know that due to Bertrand's theorem \cite{Bertrand1873} only two potentials are such as all bounded orbit are closed and the motion is periodic, namely the harmonic oscillator potential $V(r)=\frac{1}{2} k r^2$  and the Kepler potential $V(r)=-k/r$. Our idea, is to begin our analysis with the simplest yet highly nontrivial case, the harmonic oscillator potential coupled to a rotor.
\section{The Harmonic Potential: degrees of freedom and resonance} 
\label{sec3}
\noindent In the case of the harmonic potential, we deal with two translational ($x,y$) and one rotational ($\theta$) generalized coordinates, which means we have a total number of degrees of freedom given by three. For example, if we consider the case of a rod AB of lenght $L$ subject to a harmonic potential $V(x,y)=\frac{1}{2}M \omega^2(x^2+y^2)$ with $\omega:=\sqrt{\frac{k}{M}}$ with $k, M>0$, then $I=\frac{1}{12}{ML^2}$, and the Hamiltonian reads:
\begin{equation}
	H = \frac{p_x^2+p_y^2}{2M} + \frac{1}{2} M \omega^2(x^2+y^2) + \frac{6p_\theta^2}{M L^2}.
	\label{eq:hamk} 
\end{equation}
The equations of motion reads:
\begin{align}
	\ddot{x}+\omega^2 x=0\, , \qquad \ddot{y}+\omega^2 y=0 \, ,\qquad  \ddot{\theta}=0 \, ,
\end{align}
with solutions:
\begin{equation}
	x(t)=x_0 \cos (\omega t) +\frac{p_{x,0}}{M \omega}\sin(\omega t)\, ,\quad  y(t)= y_0 \cos (\omega t) +\frac{p_{y,0}}{M \omega}\sin(\omega t)\, , \quad  \theta(t)=\theta_0+\frac{p_\theta}{I} t \, ,
\end{equation}
where we defined $\omega:=\sqrt{\frac{k}{M}}$, \quad $x_0:=x(0)$, \quad $p_{x,0}:=p_x(0)$, \quad $y_0:=y(0)$, \quad $p_{y,0}:=p_y(0)$.
\begin{figure}[h!]
	\centering
	\includegraphics[width=0.4\textwidth]{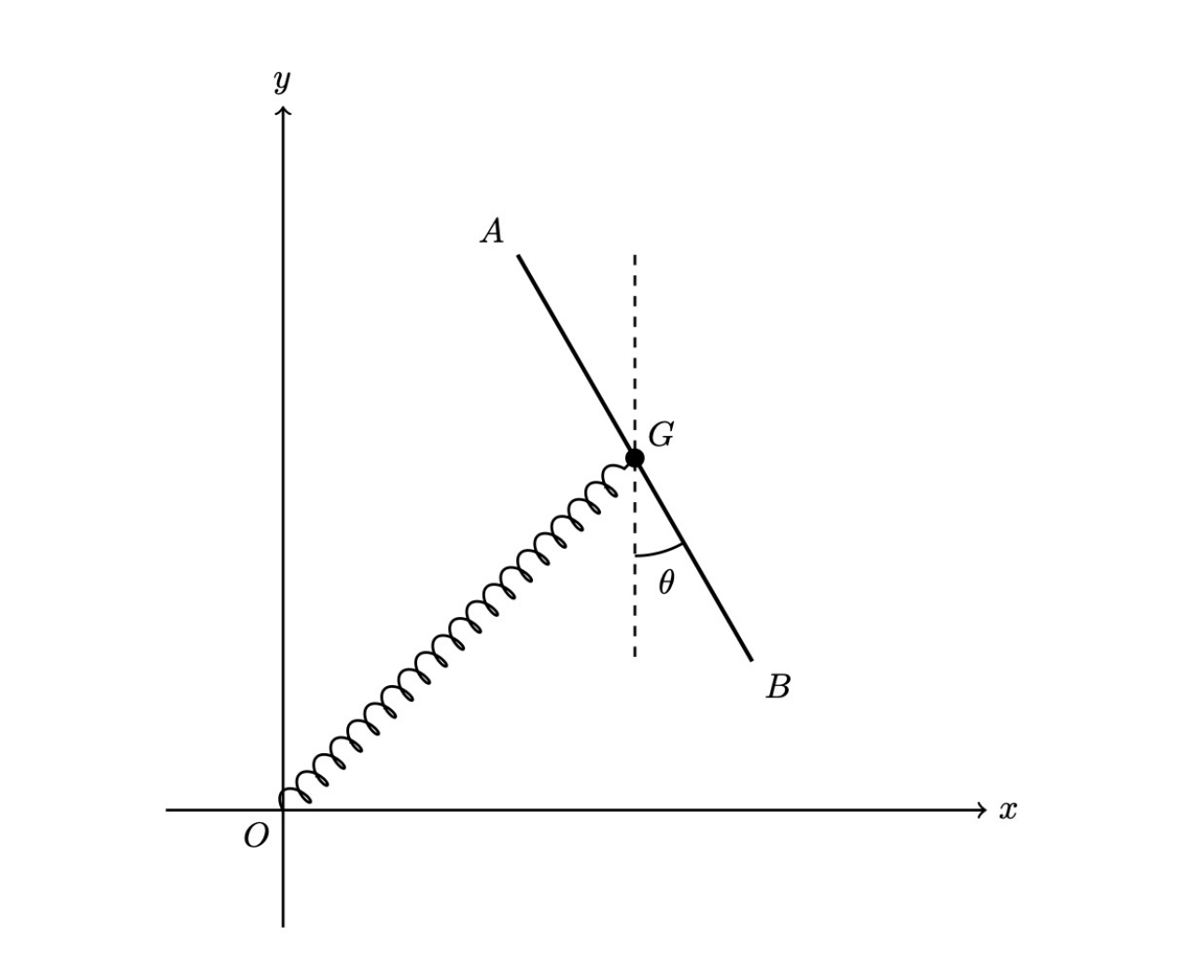}
	\includegraphics[width=0.4\textwidth]{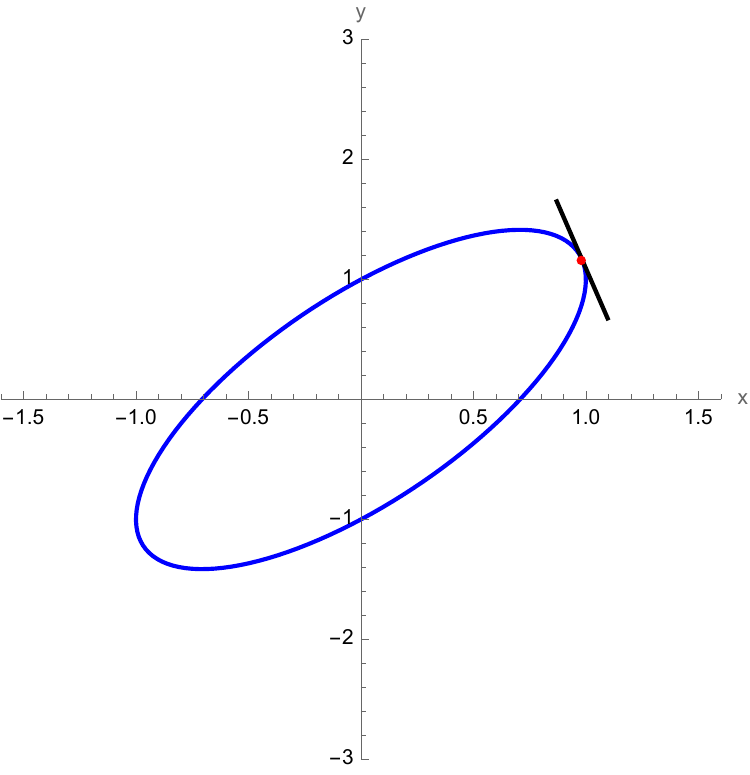}
	\caption{On the left it is reported a schematic realization of the physical problem described by the Hamiltonian \eqref{eq:hamk}, whereas on the right a schematic representation of an orbit in the plane of the same system is provided. Here, we have taken $M=1$, $k=1$ and $L=1$. }
	\label{HO}
\end{figure}

\noindent As we can see from \figref{HO} above, the orbital motion spanned by the center of mass generates a closed path (an ellipse), and this is not difficult to expect as the additional term in the Hamiltonian just represent an additive constant, being $p_\theta$ a conserved quantity ($\theta$ is a cyclic coordinate), and we know that the $2D$ isotropic oscillator is a superintegrable system. So, because of this phenomenon, the system can be reduced to an effective Hamiltonian depending on $(x,y)$ (the orbital motion is decoupled from the “internal" rotation).

\noindent However, what we are interested to understand in this work is the full dynamics, orbital plus rotational. This is because in principle, a Hamiltonian system with three degrees of freedom is characterized by a total number of five functionally independent constants of motion. Even more, if we are able to find these constants of motion, as we have a clear reduction to the standard 2D classical systems in the limit $p_{\theta} \to 0$, it would be reasonable to expect, though this should be treated with due caution, that the corresponding symmetry algebras will contain, as a subalgebra, those associated with the original planar systems, recovered in the appropriate $p_{\theta} \to 0$ limit. This perspective also suggests that these systems could be regarded as natural sources of deformed symmetry algebras. 

In order to expect maximally superintegrability for the Hamiltonian: 
\begin{equation}
	H = \frac{p_x^2+p_y^2}{2M} + \frac{M \omega^2}{2}(x^2+y^2) + \frac{p_\theta^2}{2 I} \, ,
	\label{eq:harmk} 
\end{equation}we should reasonably expect that the orbital frequency of the motion, i.e. $\omega:=\sqrt{{k/M}}$ is tuned to resonance within the  rotor frequency, From now on, if the reader prefer, can consider in his mind that $I=\frac{1}{12}ML^2$.

\noindent In general, the system \eqref{eq:harmk} will be endowed with the following linearly independent constants of motion:
\begin{align}
	F_1:=H, \quad F_2=\frac{p_\theta^2}{2I}, \quad \text{L}:=x p_y-y p_x, \quad &G_1:=\frac{p_x p_y}{2M}+\frac{M \omega^2}{2} x y, \nonumber\\  &G_2:=\frac{1}{2M}(p_x^2-p_y^2)+\frac{M \omega^2}{2} (x^2-y^2) 
	\label{consquant}
\end{align}
and one of the involutive subsets closes in the following Abelian algebraic structure:
\begin{equation}
	\{F_1,F_2\}=\{F_1, \text{L}\}=\{F_2,\text{L}\}=0 \, .
	\label{algstr}
\end{equation}
Notice that $G_1$ is nothing but the $x$-$y$ component of the Demkov-Fradkin tensor \cite{D1959,F1965}. Here, the only non abelian part will generate the three-dimensional $\mathfrak{su}(2)$ Lie algebra that we know for the standard $2D$ harmonic  oscillator:
\begin{equation}
	\{L,G_1\}=-G_2 \, \quad \, \{L,G_2\}=4 G_1 \, \quad \, \{G_1,G_2\}=- \omega^2\text{L} \, ,
	\label{eq:su(2)ab}
\end{equation}
whose Casimir is now given by:
\begin{equation}
	4G_1^2+ G_2^2+\omega^2\text{L}^2=\left(F_1-F_2\right)^2 \, .
\end{equation}
At this point, let us consider the complex ladder functions:
\begin{equation}
	u:=x+\imath y \, , \qquad w:=p_x+\imath p_y \, , \qquad Z=w+\imath M \omega u  \, ,
\end{equation}
such that:
\begin{equation}
	\{Z,F_1\}=\imath  \omega Z \, ,
\end{equation}
which means: $Z(t)=Z(0)\exp(\imath  \omega t)$, meaning that the evolution is purely oscillatory at angular frequency $\omega$: the complex variable rotates in the complex plane with constant modulus  $|Z(t)|=|Z(0)|$. At this point, we are ready to define the complex integrals (depending on $m$ and $n$):
\begin{equation}
	K_{m,n}=K_{m,n}(x,y,\theta, p_x,p_y)=Z^m  \exp(- \imath n \theta) \, ,\qquad m,n \,\,\text{integers}
	\label{compint}
\end{equation}
which are constants of motion if and only if:
\begin{equation}
	\frac{\Omega}{\omega}=\frac{p_\theta}{I \omega}=\frac{m}{n} \in \mathbb{Q}^* \, ,
	\label{eq:res}
\end{equation}
where we introduced the quantity $\Omega:=\frac{p_\theta}{I}$. Notice that $K_{km,kn}=K_{m,n}^k$. Physically,  condition \eqref{eq:res} is highly nontrivial, since it is telling us that the system will be endowed with additional integrals of motion, and so will be maximally superintegrable, if the rotation of the rigid body will resonate with the rotation governing the orbital planar motion. As a matter of fact, if we compute the Poisson brackets we get:
\begin{equation}
	\{K_{m,n},F_1\}=\imath(m\omega-n \Omega )K_{m,n} \, .
\end{equation}
This means that, $K_{m,n}$, will be a constant of motion if and only if the system will be in resonance, see \eqref{eq:res}. Let us now consider the complex ladder functions:
\begin{equation}
	\bar{u}:=x-\imath y \, , \qquad \bar{w}:=p_x-\imath p_y \, , \qquad \bar{Z}=\bar{w}-\imath M \omega \bar{u} \, ,
\end{equation}
such that:
\begin{equation}
	\{\bar{Z},F_1\}=-\imath  \omega \bar{Z} \, ,
\end{equation}
which means: $\bar{Z}(t)=\bar{Z}(0)\exp(- \imath  \omega t)$. At this point, we are ready to define the new complex conjugated integrals:
\begin{equation}
	\bar{K}_{m,n}=\bar{K}_{m,n}(x,y,\theta, p_x,p_y)=\bar{Z}^m  \exp( \imath n \theta) \, , \qquad m,n \,\,\text{integers}
	\label{compint}
\end{equation}
which are again constants of motion if and only if \eqref{eq:res} is satisfied, being:
\begin{equation}
	\{\bar{K}_{m,n},F_1\}=-\imath( m\omega-n \Omega)\bar{K}_{m,n} \, .
\end{equation}
With the help of these integrals, we are led naturally to consider the following combinations:
\begin{equation}
	P_{m,n}:=\frac{K_{m,n}+\bar{K}_{m,n}}{2} \, , \qquad Q_{m,n}:=\frac{K_{m,n}-\bar{K}_{m,n}}{2 \imath} 	\, ,
	\label{eq:eqss}
\end{equation}
for which the additional functional relations hold:
\begin{equation}
	Q_{m,n}^2+P_{m,n}^2= \bigl(2 M (F_1-F_2+\omega L)\bigl)^m \, .
	\label{funcreladd}
\end{equation}
Just to have an idea, the first few (linear in the momenta) integrals have the following form:
\begin{align}
	P_{1,1}&=\left(p_x-M \omega y \right) \cos(\theta)+\left(p_y+M \omega x \right) \sin(\theta)\\
	Q_{1,1}&=\left(p_y+M \omega x \right) \cos(\theta)+\left(-p_x+M \omega y \right) \sin(\theta)\, .
\end{align}
It is not difficult to see the structure and, also, to compute the others (in general higher-order) through the formulas \eqref{eq:eqss}. Let us emphasize once again that these integrals will Poisson commute with the Hamiltonian \eqref{eq:harmk} if and only if the resonance condition is satisfied. We mention that, and this is the most important point, if we compute the rank of the $5$ x $6$  Jacobian:
\begin{equation}
	\text{Rank}(J)=\text{Rank}\dfrac{\partial (F_1,  F_2, G_1,  G_2, P_{1,1})}{\partial (\boldsymbol{q},\boldsymbol{p})}=5 \, .
\end{equation}
If we remove $P_{1,1}$ the rank turn out to be $4$. Also adding the angular momentum \text{L} to the set does not change the result. Here we have a system in the plane that, due to the internal rotational degree of freedom, is characterized by three total degrees of freedom and then five functionally independent constants, thereby confirming its maximal superintegrability. Let us also remark once again that increasing the index $m$ would increase the order of the corresponding constant of motion. The problem of characterizing the complete symmetry algebra for such kind of systems remains, in my opinion, a (very interesting) open problem. Notice, in fact, that in general here we have a “tower'' of integrals which may be all involved to reach final closure.

 \section{Oscillator in a Vertical Plane with Gravity}  \label{sec4}
\noindent Let us consider, the same oscillator, but now in a vertical plane, so that also gravity will play a role. We have:

\begin{equation}
	H = \frac{p_x^2+p_y^2}{2M} + \frac{M \omega^2}{2}(x^2+y^2) + \frac{p_\theta^2}{2 I} +Mgy \, .
	\label{eq:harmkgfa}
\end{equation}
Assume $y$ is the vertical coordinate measured upwards. Let us define the shift coordinate:
\begin{equation}
	y'=y+\frac{g}{\omega^2} \, .
\end{equation}
Then,  considering the new equilibrium at $y=-\frac{g}{\omega^2}$, we have:
\begin{equation}
	\frac{M \omega^2}{2}(x^2+y^2) + M g y=\frac{M \omega^2}{2}\left(x^2+\left(y'-\frac{g}{\omega^2} \right)^2\right) +M g \left(y'-\frac{g}{\omega^2}\right)
	\label{eq:equili}
\end{equation}
which turns out to be:
\begin{equation}
	\frac{M \omega^2}{2}(x^2+y^2) + M g y=\frac{M \omega^2}{2}\left(x^2 +y'^2\right) -\frac{M g^2}{2 \omega^2}\, .
	\label{eq:equiliw}
\end{equation}
So that the complete Hamiltonian is: 
\begin{equation}
	H=\frac{p_x^2+p_y^2}{2M} +\frac{M \omega^2}{2}\left(x^2 +y'^2\right) -\frac{M g^2}{2 \omega^2}+ \frac{p_\theta^2}{2 I} \, .
	\label{eq:compham}
\end{equation}
This is the same isotropic oscillator in the presence of rotor,  just centered at 
$(x,y')=(0,0)$, plus a constant energy shift. Now, the question naturally arises, what about the $\mathfrak{su}(2)$ symmetry when the rotational invariance is broken.  Again, we can consider the following (standard, i.e. $(x,y)$) quantities and define the set \eqref{consquant}, where $F_1$ is given by $H$ \eqref{eq:harmkgfa}. Here the set $\{\text{L}, G_1, G_2\}$ should span the $\mathfrak{su}(2)$ algebra in absence of gravity. However, it is clear that, in presence of gravity \text{L} is no more a conserved quantity as we get:
\begin{equation}
	\{F_1, \text{L}\}=M g x \neq 0 \, ,
\end{equation}
which is zero, only if $g=0$. However, if one recenter at the shifted origin (use $y'$)  then the “correct” angular momentum about the equilibrium point is:
\begin{equation}
	\text{L}'=x p_y-y' p_x=\text{L}-\frac{g}{\omega^2}p_x
	\label{eq:correequil}
\end{equation}
which Poisson commutes with the Hamiltonian:
\begin{equation}
	\{F_1, \text{L}'\}=0 \, .
\end{equation}
The same is true for $F_2$ of course. In the same way, one should redefine the quadratic generators with  $y'$ in such a way to get:
\begin{align}
	G_1'&:=\frac{p_x p_y}{2M}+\frac{M \omega^2}{2} x y' =\frac{p_x p_y}{2M}+\frac{M \omega^2}{2} x y +\frac{M g x}{2}\\ G_2'&:=\frac{1}{2M}(p_x^2-p_y^2)+\frac{M \omega^2}{2} (x^2-y'^2)\nonumber\\&\,\,=\frac{1}{2M}(p_x^2-p_y^2)+\frac{M \omega^2}{2} (x^2-y^2) -M g y -\frac{Mg^2}{2\omega^2} \, ,
\end{align}
and, consequently:
\begin{equation}
	\{F_1, G_1'\}= \{F_1, G_2'\}=0 \, ,
\end{equation}
with the symmetry algebra being:
\begin{equation}
	\{L',G_1'\}=-G_2' \, \quad \, \{L',G_2'\}=4 G_1' \, \quad \, \{G_1',G_2'\}=- \omega^2\text{L}' \, ,
	\label{eq:su(2)ab}
\end{equation}
where the Casimir is now given by:
\begin{equation}
	4G_1'^2+ G_2'^2+\omega^2\text{L}'^2=\left(F_1-F_2+\frac{M g^2}{2 \omega^2}\right)^2 \, .
\end{equation}
It is clear that, also in this case, since the only difference with respect to the previous case is a constant shift, the system will admit analogous additional integrals of motion. The only physical change is that the rotation is now taken about the point $\frac{g}{\omega^2}$.

\section{Conclusions and open problems} 
\label{sec5}
	\noindent We have introduced a systematic way to construct \emph{extended} planar superintegrable models by coupling a standard two--dimensional Hamiltonian system to a rigid-body rotor, thereby producing systems with three degrees of freedom (two translational and one rotational). The central mechanism is the \emph{resonance condition} between the orbital motion in the plane and the internal rotation: when the corresponding frequencies are commensurate, additional (generically higher-order) integrals of motion emerge. In this sense, resonance plays the role of a ``superintegrability selector'', distinguishing regimes in which the extended dynamics admits new conserved quantities and a richer algebraic structure.

\begin{figure}[h!]
	\centering
	\includegraphics[width=0.4\textwidth]{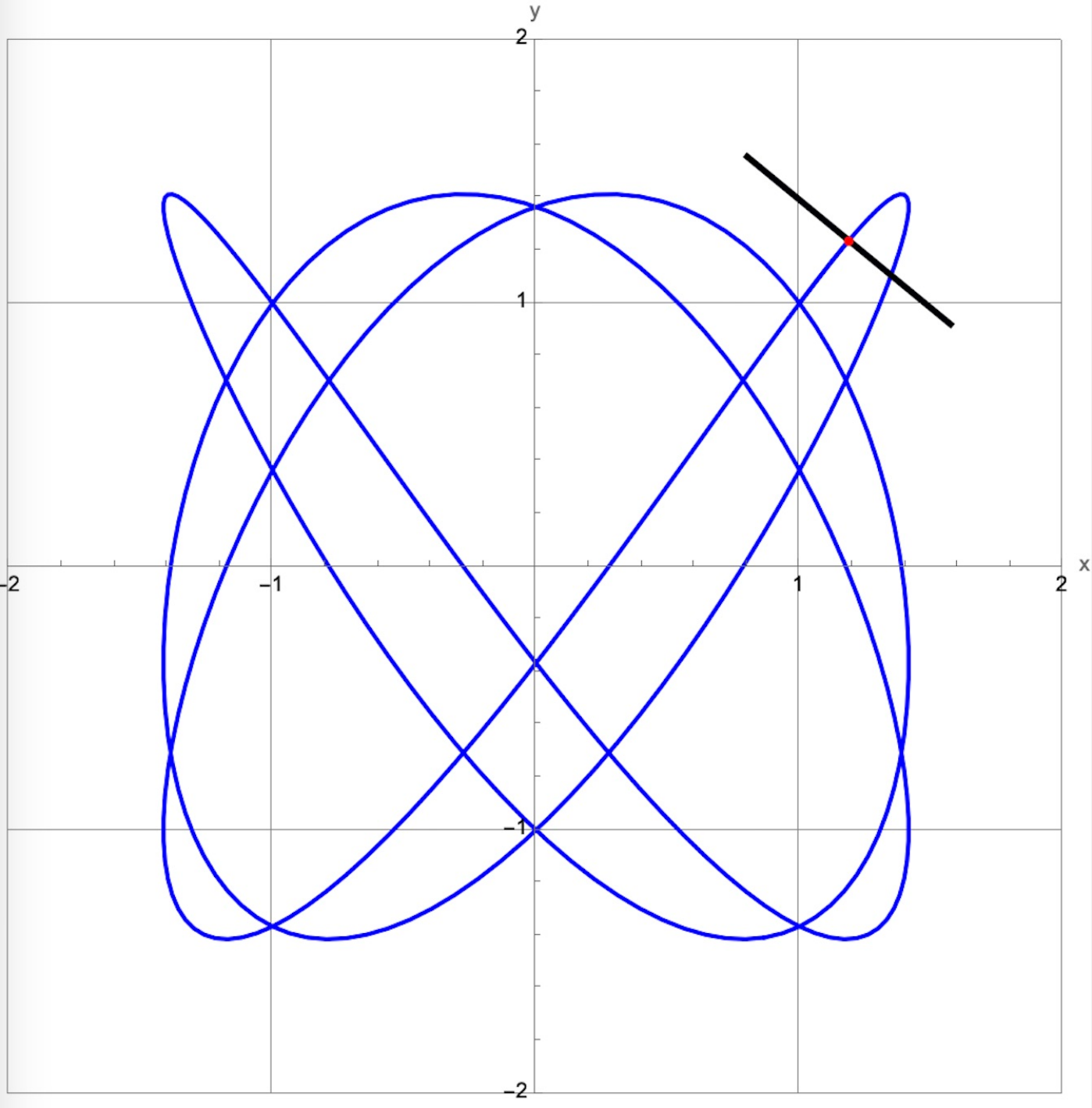}
	\caption{Schematic representation of the physical trajectory described by the Hamiltonian of an anisotropic oscillator, with frequency ratio $\omega_x:\omega_y=3:5$, coupled to a rotor. Here we still consider a homogeneous rod with mass $M=1$ and length $L=1$.}
	\label{AHO}
\end{figure}

The rotor coupling naturally enlarges the symmetry content of the underlying planar problem. In the harmonic case the familiar $\mathfrak{su}(2)$ structure would survive as a subalgebra and is expected to be extended by the resonant integrals. Ideally, our analysis would also fit within the broader context of the four classical flat second-order superintegrable systems in $N=2$. Here we report, as an higher order example, the anisotropic oscillator, for which we numerically get something as reported in \figref{AHO}.

\noindent A guiding expectation is that \emph{maximal} superintegrability of the three-degree-of-freedom extension should manifest itself through the presence of $2N-1=5$ functionally independent integrals of motion. The resonance-induced constants discussed here provide a concrete route to achieving such maximality, and they naturally point toward higher-order generalizations, in close analogy with the anisotropic oscillator where commensurability is responsible for additional integrals.

Several directions appear particularly promising. The two-dimensional potential cases are already highly interesting in their own right, for a variety of dynamical and algebraic reasons. Moreover, allowing for an additional angular potential $V=V(\theta)$ could lead to genuinely new phenomena. More generally, the quantization of these translational–rotational models should clarify how the resonance condition manifests itself at the spectral level, and how the enlarged symmetry algebras act on the quantum Hilbert space, once the relevant representation-theoretic framework is identified, Hamiltonian included. It is also unclear to the author whether these models could find applications in more exotic settings, such as satellite dynamics.
A second natural direction is to extend the construction to higher dimensions, coupling 
$N$ translational degrees of freedom to one (or more) internal rotors, and to investigate how resonance organizes families of superintegrable systems and their polynomial symmetry algebras. More broadly, we expect rotor-extended models to provide a flexible framework for generating new superintegrable Hamiltonians whose integrals of motion and symmetry structures are not apparent in purely translational settings, but a full characterization is expected to be highly nontrivial, and likely enormously complex.
\subsection*{Declaration of generative AI and AI-assisted technologies in the writing process}
\noindent During the preparation of this work, the author used ChatGPT by OpenAI in order to identify grammatical errors and to aid in the overall readability of the text. After using this tool, the output was reviewed and edited as needed and the author takes full responsibility for the content of the publication.

\subsection*{Acknowledgements}

The author thanks the referee for carefully reading the manuscript and for providing valuable feedback. D.L. is supported by HORIZON EUROPE - European Research Council (ERC) - STARTING GRANT 2021 “Hamiltonian Dynamics, Normal Forms and Water Waves” (HamDyWWa) - Project Number: HE$\_$ERC22RMONT$\_$01.
Views and opinions expressed are however those of the author only and do not necessarily reflect those
of the European Union or the European Research Council. Neither the European Union nor the granting authority can be held responsible for them. He is also partially supported by INFN-CSN4 (Commissione Scientifica Nazionale 4 - Fisica Teorica), MMNLP project. D.L. is a member of GNFM, INdAM. The research of D.L, at the time the paper was written, has been also partially funded by MUR - Dipartimento di Eccellenza 2023-2027, codice CUP G43C22004580005 - codice progetto   DECC23$\_$012$\_$DIP.

\label{lastpage}

\begin{thebibliography}{99}


\bibitem{MPW}
W. Jr. Miller, S. Post, P. Winternitz,  Classical and quantum superintegrability with applications, \href{https://doi.org/10.1088/1751-8113/46/42/423001}{J. Phys. A: Math. Theor. \textbf{46} (2013) 423001}.


\bibitem{BBM}
I. A. Bizyaev, A. V. Borisov, I. S. Mamaev, Superintegrable generalizations of the Kepler and Hook problems, \href{https://doi.org/10.1134/S1560354714030095}{Regular Chaot. Dyn. \textbf{19} (2014) 415}.

\bibitem{Liouville1853}
J. Liouville, Note sur l'int\'egration des \'equations diff\'erentielles de la Dynamique, pr\'esent\'ee au Bureau des Longitudes le 29 juin 1853 \href{http://www.numdam.org/item/JMPA_1855_1_20__137_0/}{J. Math. Pures Appl. \textbf{20} (1855) 137}.

\bibitem{whittaker}
E. T. Whittaker, A Treatise on the Analytical Dynamics of Particles and Rigid Bodies, Cambridge University Press, Cambridge, 1999.

\bibitem{Arnol'd1997}
V. I. Arnol'd, Mathematical Methods of Classical Mechanics 2nd ed, Graduate Texts in Mathematics   \textbf{60},  Springer,  Berlin, 1997.

\bibitem{Reshetikhin}
N. Reshetikhin, Degenerately integrable systems, \href{https://doi.org/10.1007/s10958-016-2738-9}{J. Math. Sci. \textbf{213} (2016) 769}.

\bibitem{Fasso}
F. Fass\`o, Superintegrable Hamiltonian systems: Geometry and perturbations, \href{https://doi.org/10.1007/s10440-005-1139-8}{Acta Appl. Math. \textbf{87} (2005) 93}.

\bibitem{10.1088/978-0-7503-1314-8}
E. G. Kalnins, J. M. Kress, and W.Miller, Separation of Variables and Superintegrability, \href{https://iopscience.iop.org/book/mono/978-0-7503-1314-8}{IOP Publishing (2018), 2053-2563}


\bibitem{FMSUW}
J. Fri$\check{s}$, V. Mandrosov, Ya.A. Smorodinsky, M.~Uhl\'i\v{r}, and
P. Winternitz, On higher symmetries in quantum mechanics, \href{https://doi.org/10.1016/0031-9163(65)90885-1}{Phys. Lett. {\bf 16}  (1965) 354}.

\bibitem{WSUF} 
P. Winternitz, Ya.A. Smorodinsky, M.~Uhl\'i\v{r}, and J. Fri$\check{s}$, Symmetry Groups In Classical and Quantum Mechanics (in Russian) \href{https://www.osti.gov/biblio/4512521}{Yad. Fiz. {\bf 4}}  
(1966) 625. Sov. Journ. Nucl. Phys {\bf 4} (1967) 444 (English Translation)

\bibitem{MSVW}
A. A. Makarov, J. A. Smorodinsky, kh. Valiev and P. Winternitz, A systematic search for nonrelativistic systems with dynamical symmetries. \href{https://doi.org/10.1007/BF02755212}{Nuovo Cimento A  {\bf 52} (1965-1970), 1061–1084 (1967)}. 

\bibitem{KSV}
J. Kress, K. Schöbel, A. Vollmer, An Algebraic Geometric Foundation for a Classification of Second-Order Superintegrable Systems in Arbitrary Dimension,  \href{https://doi.org/10.1007/s12220-023-01413-8
}{J. Geom. Anal. {\bf 33} (2023) 360}.

\bibitem{AS}
NIST Digital Library of Mathematical Functions, \href{https://dlmf.nist.gov//}{National Institute of Standards and Technology}. 

\bibitem{TTW}
P. Tempesta, A. V. Turbiner, P. Winternitz; Exact solvability of superintegrable systems. J. Math. Phys. {\bf 42} )(2001) 4248–4257

\bibitem{Bertrand1873}
J. Bertrand, Th\'eor\`eme relatif au mouvement d'un point attir\'e vers un centre fixe, C. R. Acad. Sci. \textbf{77} (1873) 849.
%
\bibitem {D1959}
Y. N. Demkov, Symmetry group of the isotropic oscillator, \href{http://jetp.ras.ru/cgi-bin/e/index/e/9/1/p63?a=list}{J. Exp. Theor. Phys. \textbf{9} (1959) 63}.
%
\bibitem{F1965}
D. M. Fradkin, Three-dimensional isotropic harmonic oscillator and ${\mathrm{SU}}_3$, \href{https://doi.org/10.1119/1.1971373}{Am. J. Phys. \textbf{33} (1965) 207}.

\end{thebibliography}
\end{document}